\documentstyle[12pt]{article}

\begin{document}

\begin{flushright}
Preprint CAMTP/94-10\\
November 1994\\
\end{flushright}
\begin{center}
\large
{\bf  Separating the regular and irregular energy levels and  their statistics
in Hamiltonian system with mixed classical dynamics}\\
\vspace{0.3in}
\normalsize
Baowen Li\footnote{e-mail Baowen.Li@UNI-MB.SI}
 and Marko Robnik\footnote{e-mail Robnik@UNI-MB.SI}\\
\vspace{0.2in} Center for Applied Mathematics and Theoretical Physics,\\
University of Maribor, Krekova 2, SLO-62000 Maribor, Slovenia\\
\end{center}

\vspace{0.4in}
{\bf Abstract.}  We look at the
high-lying eigenstates (from the 10,001st to the 13,000th) in the Robnik
billiard
(defined as a quadratic conformal map of the unit disk) with the shape
parameter $\lambda=0.15$. All the 3,000 eigenstates have been numerically
calculated and examined in the configuration space  and in the phase space
which - in comparison with the classical phase space - enabled a clear cut
classification of energy levels into regular and irregular. This is the first
successful separation of energy levels based on purely  dynamical rather than
special geometrical symmetry properties. We calculate the fractional measure of
regular levels as $\rho_1=0.365\pm 0.01$ which is in remarkable agreement with
the classical estimate $\rho_1=0.360\pm 0.001$. This finding confirms the
Percival's (1973)  classification scheme, the assumption in Berry-Robnik (1984)
theory and the rigorous result by Lazutkin (1981,1991).
The regular levels obey the Poissonian
statistics quite well whereas the irregular sequence exhibits the fractional
power law level repulsion and globally Brody-like statistics with $\beta =
0.286\pm0.001$.  This is due to the strong localization of irregular
eigenstates
in the classically chaotic regions. Therefore in the entire spectrum we see
that  the Berry-Robnik regime is not yet fully established so that the level
spacing distribution is correctly captured by the Berry-Robnik-Brody
distribution (Prosen and Robnik 1994).
\\\\

\\\\
Submitted to {\bf Journal of Physics A}

\normalsize
\vspace{0.3in}
\newpage

\section{Introduction}

In the early days of quantum chaos Percival (1973) has suggested,
using the semiclassical picture and the correspondence principle, that the
eigenstates of a classically nonintegrable and (partially) chaotic Hamiltonian
system should be classified as regular or irregular, depending on whether
they are associated with a classically regular or chaotic regions in the phase
space.  In fact, he referred mainly to the sensitivity of the
energy levels with respect to some family parameter (of a one-parameter family
of Hamiltonians), by saying that in irregular levels the second differences
(now known and studied as curvature) are typically much larger than in regular
levels. This picture has been recently confirmed in the studies of curvature
distribution of quantum spectra of classically nonintegrable systems (Gaspard
{\em et al} 1990 , Takami and Hasegawa 1992)
where it has been finally demonstrated
that here again we encounter universality classes (Zakrzewski and Delande 1993,
von Oppen 1994a,b).
\\\\
On the other hand the irregular levels have been supposed to be associated with
the classically chaotic regions also geometrically in the sense of the
Principle of Uniform Semiclassical Condensation (PUSC, see e.g. Li and Robnik
1994a, Berry 1977a,b), which states that the quantal phase space distribution
function like Wigner or Husimi or anything in between in the semiclassical
limit
$\hbar\rightarrow 0$ uniformly condenses on the associated classical invariant
object, which in this case is the underlying chaotic component. According to
the same principle the regular levels are associated with the quantized
invariant tori. Thus in the semiclassical limit the entire spectrum and the
associated eigenstates are decomposed in the regular sequence and generally
many irregular sequences. More quantatively, it is intuitively obvious to
assume that the fractional level density of each level sequence is equal to the
fractional phase space volume of the underlying invariant classical component,
and this is one of the main assumptions of the Berry-Robnik (1984) theory. It
is also a rigorous result by Lazutkin (1981, 1991) that the torus quantization
in the plane classical convex billiards works only for states where the
classical invariant tori exist and the relative measure of such regular states,
according to him, is equal to the fractional phase space volume of the
classically regular regions.
\\\\
So far an inspection and separation of spectra in
this sense has been performed only in one special case (Bohigas, Tomsovic
and  Ullmo 1990a-b, 1993), where the regular levels are characterized by being
almost degenerate pairs due to some discrete geometrical symmetry. It is the
goal of our present paper to perform a separation of regular and irregular
eigenstates on purely dynamical grounds in a generic system of mixed type
classical dynamics described by a KAM-type scenario, which is to the best of
our knowledge the first such analysis. The system that we shall analyze
numerically is the Robnik billiard (1983,1984) (defined as a quadratic
conformal map of the unit disk) with the shape parameter $\lambda=0.15$, and
this work is a natural continuation of our previous work (Li and Robnik 1994b)
\\\\
We have calculated the 3,000 eigenstates between the 10,001st and the
13,000th of even
parity in configuration space and in the phase space and performed the
classification of states in comparison with the classical dynamics (Poincar\'e
surface of section plots). After this
separation of regular and irregular states we have done rather complete
spectral statistical analysis which we present in section 3. As we shall see
the ingredients of the Berry-Robnik picture are almost completely confirmed,
except for the localization phenomena in the chaotic states which are
responsible for the deviation from GOE statistics to which ideally the
statistics must converge in the strict semiclassical limit $\hbar\rightarrow0$.
These phenomena have been recently demonstrated, discussed and partially
(qualitatively) explained by Prosen and Robnik (1994b).

\section{The preliminaries: The calculational technique and the method of
analysis}

The theoretical questions that we addressed in the introduction are of course
completely general but in order to illustrate them we have to confine ourselves
to some specific and possibly generic system. For such a purpose it is ideal to
study billiard systems. We have chosen the Robnik billiard (defined as a
quadratic conformal map of the unit disk $w=z+\lambda z^2$) with the  shape
parameter $\lambda=0.15$, which is a convex plane billiard. This system and the
conformal mapping diagonalization technique has been introduced by  Robnik
(1983, 1984) and further studied by Berry and Robnik (1986), Robnik and Berry
(1986),  Hayli {\em et al}
(1987), Frisk (1990), Bruus and Stone (1994), Stone and Bruus (1993,1994),
Markarian (1993), Prosen and Robnik (1993,1994b), Li and Robnik (1994a,b) and
B\"acker and Steiner (1994). This one parameter system (called Robnik billiard
or Robnik model by most workers in the field) is a nice generic system which at
$\lambda=0$ is the integrable circle billiard, for $0<\lambda \le 1/4$ it is a
typical KAM system (the KAM theorem applies because the boundary is analytic
and convex), for $\lambda>1/4$ it becomes non-convex and eventually for some
sufficiently large $\lambda\le 1/2$ becomes ergodic, mixing and K-system.
Markarian (1993) has recently rigorously proved this property for
$\lambda=1/2$ (cardiod billiard). Li and Robnik (1994c) have numerical evidence
that ergodicity (and mixing and K property) exists already at $\lambda \ge
0.2775$, but could set in even earlier. More details about the system and the
numerical technique are given in our recent papers (Li and Robnik 1994a,b).
\\\\
In calculating the eigenfunctions in configuration space we have used the
Heller's method (1991) of plane wave decomposition by applying the singular
value decomposition (Press {\em et al} 1986). All the technical details and
tricks have been described in (Li and Robnik 1994a,b). The problem with missing
eigenstates has been overcome by using the exact (double precision: 16 digits)
eigenenergies for all 3,000 consecutive states, from the 10,001st to the
13,000th
state. The exact energy levels have been obtained by the conformal mapping
diagonalization technique described in detail in (Prosen and Robnik 1993a).
\\\\
We have calculated the eigenfunctions in configuration space, their
Wigner functions in phase space, also their projections onto the surface of
section,  and finally also their Husimi type smoothed objects on the surface of
section. All the definitions, formulae and the method of presentation are
exactly as introduced in (Prosen and Robnik 1993b), and later employed in (Li
and Robnik 1994b). Therefore and because we shall not show any plots of
eigenfunctions in this
paper we refer the reader to those previous works.  The quantal phase space
plots (smoothed projections of Wigner functions onto SOS) have been compared
with the classical SOS plots for all the 3,000 states. (Of course we have also
plotted the configurational eigenfunctions but this is not so important for the
classification of states.) One smaller but representative part of this output
(200 eigenstates) shall be published separately (Li and Robnik 1994d)
and is also available from the
authors upon a request. Some representative eigenstates are shown in (Li and
Robnik 1994b).
\\\\
In spite of the vast amount of our material
(altogether 2$\times$3,000 plots) there was no difficulty in classifying the
eigenstates. A state is declared regular if it is localized on invariant torus
belonging to the classical regular region in the classical SOS. A state is
declared irregular if it is localized (either uniformly or non-uniformly) on a
classically chaotic region in the SOS.  According to the PUSC we expect that in
the strict semiclassical limit $\hbar\rightarrow 0$, or equivalently when the
energy goes to infinity, the irregular states become uniformly extended on the
classical chaotic component. However, before this limit is reached we observe
strong localization of irregular states due to the slow classical diffusion
inside the underlying classical chaotic component.  In fact the vast majority
of our irregular states are localized chaotic states and only a few irregular
states are extended chaotic. The consequences of such localization for the
spectral statistics have been recently discussed and demonstrated by Prosen and
Robnik (1994b): For the associated irregular level sequence we find the
phenomenon of the fractional power law level repulsion and globally a
Brody-like behaviour which then approaches GOE deeper in the semiclassical
limit of smaller and smaller effective $\hbar$.  This is confirmed again in
our present paper in section 3.
\\\\
In performing the classification of states we had 109 cases which at first
sight would be classified as mixed states in the sense that they "live" both
partially in classically regular regions and partially in irregular regions.
However a closer examination of finer details of the quantal phase space
functions (by decreasing the smoothing length) led us to the conclusion that
they should be classified as regular states. Other a posteriori reasons for
such decision will be described in the next section.
\\\\
The central result of this paper is the counting of regular and irregular
states. Among 3,000 states we found 987 regular, 109 mixed and 1904 irregular
states. The percentages are 32.9\%, 3.6\% and 63.5\%, respectively. This is the
result of the preliminary classification. With the decision mentioned before
and further explained in section 3 we absorb the mixed states in the set of
regular states (so now there are 987+109 =1096 of them)
which results then in $\rho_1=0.365 \pm 0.01$ for the regular
states and $\rho_2=1-\rho_1=0.635\pm0.01$ for irregular states.  This is in
excellent agreement with the classical relative phase space volume of the
regular regions as calculated and reported by Prosen and Robnik (1993a) where
$\rho_1=0.360\pm0.001$. We believe that this is quite striking confirmation not
only of Percival's scheme (1973), but also of the main assumption in
Berry-Robnik (1984) theory of energy level spacings and more specifically of
Lazutkin's rigorous results (1981,1991) on convex plane billiards.
\\\\
In order to get some idea about the convergence of these numerical figures with
the size of the sample we should note that examination of additional 1,000
consecutive states, namely from the 13,001st to the 14,000th state, and the
classification by the same criteria as above gives the result $\rho_1 =
0.360\pm0.01$ which is then in brilliant agreement with the classical value.
However, due to the statistical fluctuations all the statistical spectral
measures do not necessarily improve monotonically as we shall comment in
the next section.
\\\\
We believe that this is the first detailed dynamical analysis of a large sample
of eigenstates enabling the separation of regular and irregular states based on
the classification procedure explained above. This successful separation of
regular and irregular states makes it possible to analyze the statistical
properties of regular and irregular sequences separately with the goal to
confirm the aspects of the Berry-Robnik approach (1984) which is the
subject of the next section.

\section{Statistics of energy level sequences}

The subject of the Berry-Robnik (1984) theory is the statistics of energy
spectra
of quantal Hamiltonians whose classical counterparts have mixed classical
dynamics in the sense that classical regular regions and classical irregular
regions coexist in the phase space. It is based on our knowledge of spectral
fluctuations and their statistics in the context of quantum chaos (Berry 1983,
Giannoni {\em et al} 1991, Haake 1991, Gutzwiller 1990, Eckhardt 1988, Bohigas
and Giannoni 1991, and Robnik 1994).
We know that there are three universality classes of
spectral fluctuations: Poisson statistics in the classically integrable
cases; in case of classical ergodicity we find the GOE/GUE statistics of
random matrix theories depending on whether there is one/none antiunitary
symmetry (we ignore spin). The interesting and difficult case of mixed
type classical dynamics of KAM-like (generic) systems has been studied for
the first time by Robnik (1984) numerically, where a continuous transition
from Poisson to GOE statistics in a billiard system (Robnik 1983) has been
found, and this work has been substantially revised in
(Prosen and Robnik 1993a).
Further theoretical progress was published by Berry and Robnik (1984)
where the following semiclassical theory of the level spacings has been
presented. The eigenstates (their Wigner functions in phase space) are
supposed to condense uniformly on the underlying classical invariant
regions such that each of them --- in the semiclassical limit --- supports
a level sequence which for itself has a Poisson or GOE statistics if the
region is regular or irregular, respectively. All the regular regions
can be thought of as supporting a single Poisson sequence because
the Poisson statistics is preserved upon a statistically independent
superposition. The mean level spacing of such a sequence is determined
by the fractional phase space volume of the regular regions.
On the other hand each chaotic (GOE) level sequence has a mean level
spacing governed by the corresponding fractional phase space volume.
The entire spectrum is then assumed to be a statistically independent
superposition of all subsequences. The statistical independence in the
semiclassical limit is justified by the principle of uniform semiclassical
condensation of eigenstates (in the phase space) and by the lack of their
mutual overlap, in consistency with Percival's (1973) conjecture.
Thus the problem of the statistics of the entire spectrum is now
mathematically precisely formulated (this forms the essence of the
Berry-Robnik approach) and its solution as far as the
level spacings are concerned can be expressed in the following way: The
statistical independence of superposition implies factorization of the gap
distribution functions (Mehta 1991, Haake 1991): The probability that there
is no level within a gap clearly factorizes upon a statistically
independent superposition. The connection between the level spacing
distribution $P(S)$ and the gap distribution $E(S)$ is as follows
\begin{equation}
P(S) = \frac{d^2 E(S)}{dS^2} \label{eq:PfromE}
\end{equation}
and conversely
\begin{equation}
E(S) = \int_{S}^\infty dx (x - S) P(x). \label{eq:EfromP}
\end{equation}
Leaving aside the general case of infinitely many chaotic components
which does not include anything surprisingly new let us restrict to the
case of one regular component with mean level density $\rho_1$ (= fractional
phase space volume) and
one chaotic component with the mean level density $\rho_2$ where
$\rho_1 + \rho_2 = 1$. This is going to be already an excellent approximation
because in a generic system of a mixed type there is usually only one large
and dominating chaotic region. Following (Mehta 1991, Haake 1991 and
Berry and Robnik 1984) we have
\begin{equation}
E(S) = E_{\rm Poisson}(\rho_1 S) E_{\rm GOE}(\rho_2 S)
\label{eq:EBR}
\end{equation}
where the Poissonian gap distribution $E_{\rm Poisson}$ is
\begin{equation}
E_{\rm Poisson}(S) = \exp(-S) \label{eq:EPoisson}
\end{equation}
whereas for the $E_{\rm GOE}$ there is no simple closed formula
(for the infinitely dimensional GOE case) and must be worked out by using
practical approximations for $P_{\rm GOE}$ and/or $E_{\rm GOE}$ which e.g.
can be found in (Haake 1991, pp72-74). However 2-dim GOE case (the so called
Wigner surmise) can be worked out explicitly as given in
(Berry and Robnik 1984, formula (28)) which usually is a good starting
approximation.

As for the delta statistics $\Delta(L)$ the similar procedure based on the
assumption of statistical independence leads to the simple (additive) formula
(Seligman and Verbaarschot 1985)
\begin{equation}
\Delta(L) = \Delta_{\rm Poisson}(\rho_1 L) + \Delta_{\rm GOE}(\rho_2 L)
\label{eq:DSV2}
\end{equation}
where $\Delta_{\rm Poisson}(L) = L/15$ whilst for $\Delta_{\rm GOE}$
there are good approximations given in (Bohigas 1991).
\\\\
The main objective of this paper is to verify the elements and aspects of
the Berry-Robnik approach.
However, before such a regime is formed in a KAM system in the
ultimate far semiclassical limit we typically observe a quasi-universal
behaviour in the spectral statistics which is characterized by the
fractional power law level repulsion and globally the adequacy of Brody
(1973,1981) distribution and of similar distributions such as Izrailev (1989).
A thorough numerical study of this phenomenon has been published recently
by Prosen and Robnik (1993a, 1994a,b), and will be discussed later on.
In cases where the chaotic component becomes large ($\rho_2\approx 1$) and the
quantal chaotic states strongly localized the phenomenological
Berry-Robnik-Brody distribution proposed in (Prosen and Robnik 1994b) can be
very efficient in capturing the global features of the experimental and
numerical data (Lopac {\em et al} 1992, 1994).
\\\\
As explained in section 2 we have separated  1096 regular levels from the
remaining 1904 irregular levels. Now we perform the statistical analysis of
each subsequence in the next two subsections 3.1-3.2 and then also of the
entire
spectrum in subsection 3.3.

\subsection{Regular levels}

Of course, before performing the separation procedure we have carefully
unfolded the entire spectrum of 3,000 levels by using the Weyl formula (with
perimeter and curvature corrections) for even parity states given in formula
(4) of our paper (Li and Robnik 1994a).
After extracting 1096 regular levels and calculating the (nearest neighbour)
level spacings we have first to normalize the empirical level spacing
distribution $P(S)$ to the unit first moment, i.e. such that the mean level
spacing is unity. By doing this we get again the relative level density of the
regular component with the value of $\rho_1=0.365$. Having performed this
normalization we calculate the cumulative level spacing distribution
$W(S)=\int_{0}^{S}P(x)dx$ and compare it with the Poisson statistics and also
try to fit it with the best fit Brody distribution
\begin{equation}
W^{B}(S,\beta) = 1 - \exp(-bS^{\beta+1}),\qquad  b=\{\Gamma
((\beta+2)/(\beta+1)) \}^{\beta+1},
\label{eq:WBrody}
\end{equation}
and best fit Berry-Robnik distribution (\ref{eq:EBR}). The result is shown in
figure 1, where we also show the so-called U-function introduced in (Prosen and
Robnik 1993a), defined as
\begin{equation}
U(W) = \frac{2}{\pi}\arccos\sqrt{1-W},
\label{eq:Ufunction}
\end{equation}
in reference with the best fit Berry-Robnik distribution.  As we see the
agreement of our numerical data with the Poissonian distribution is reasonably
good in the sense that Poisson is within the $\pm1\sigma$  statistical
fluctuations of the data. The best fit Brody distribution and the  best fit
Berry-Robnik distribution deviate slightly from Poisson: For Brody we get
$\beta =0.041\pm0.002$, which ideally should be zero, and for Berry-Robnik  we
get $\rho_1=0.704$, which ideally should be unity. In figure 2 we show the
delta
statistics for the same sequence of 1096  regular levels and for sufficiently
small $L<L_{max}$ (for $L$ larger than $L_{max}$ the saturation effects set in
(Berry 1985): in our case  $L_{max}\approx 10$) we see excellent agreement with
Poisson. The best fit Seligman-Verbaarschot delta statistics (\ref{eq:DSV2})
gives $\rho_1=0.9998$, which is surprisingly close to the ideal value unity.
Thus we think that our statistical analysis of the regular sequence of levels
clearly supports the theoretical assumptions in the Berry-Robnik (1984)
approach.

\subsection{Irregular levels}

In analogy to the previous procedure we have extracted 1904 irregular levels
and normalized the level spacing distribution to the unit first moment which
yields $\rho_2=0.634\pm0.002$ in consistency with $\rho_1 =0.365\pm0.002$,
such that we have unit level density of the entire spectrum,
namely $\rho_1 +\rho_2 =1$.
The cumulative level spacing distribution for the irregular sequence is
plotted in figure 3, where we see that it deviates from the ideal GOE
distribution substantially
and significantly. This is a consequence of the strong localization of
irregular eigenstates on classically chaotic components demonstrated for even
higher energies (ten times larger) in (Li and Robnik 1994b) for the same
billiard with $\lambda =0.15$, whereas the implications for level statistics
have been demonstrated and discussed by Prosen and Robnik (1994b). The best fit
Berry-Robnik distribution yields $\rho_1 =0.374$ which however ideally
should be zero. On the other hand the more interesting best fit Brody
distribution captures the fractional power law level repulsion with
$\beta =0.286\pm0.001$, and
as we see both in the $W(S)$ as well as in the U-function plot it is globally
much better fit than Berry-Robnik which is qualitatively well understood
in (Prosen and Robnik 1994b). If we were able to go to much higher energies,
say 100 times higher, then the statistics of irregular levels is predicted to
approach GOE statistics, i.e. the Brody parameter $\beta$ goes to one.
\\\\
As for the delta statistics the analysis of the data by the best fit
Seligman-Verbaarschot formula (\ref{eq:DSV2}) gives $\rho_1=0.373$, which is
equal to and consistent with the Berry-Robnik fit for level spacings, but of
course is at variance with the ideal value $\rho_1=0$. This is certainly the
consequence of the localization effects. Having understood the reasons for
the deviation of spectral statistics from the limiting (as $\hbar\rightarrow
0$ or equivalently as the energy goes to infinity) GOE statistics we may
conclude that the relevant aspect of the Berry-Robnik (1984) theory is
reconfirmed supporting our claim that this theory is asymptotically exact
theory.

\subsection{The entire spectrum}

Ideally in the strict semiclassical limit of
sufficiently high energies we would expect the statistics of the entire
spectrum of 3,000 levels (regular plus irregular) to be described by the
Berry-Robnik distribution (\ref{eq:EBR}). This prediction is based on two
facts: (i) GOE applies for irregular levels, and (ii) regular and irregular
levels are superposed statistically independently. Now, the first assumption is
not satisfied due to the localization effects explained in the preceeding
subsection, so that instead of GOE we find in fact Brody with $\beta
=0.286\pm0.001$, as shown in figure 3.
The second assumption of statistical independence is accepted as
valid and thus the reasoning leading to the factorization of the gap
distribution is valid except that we find now not Berry-Robnik distribution but
the so-called Berry-Robnik-Brody (Prosen and Robnik 1994b) distribution instead
\begin{equation}
E^{BRB}(S,\rho_1,\beta) = E_{Poisson}(\rho_{1}S)E^{B}(\rho_{2}S,\beta)
= e^{-\rho_{1}S}E^{B}((1-\rho_1))S,\beta),
\label{eq:EBRB}
\end{equation}
where Brody statistics has gap distribution which can be expressed in terms of
incomplete Gamma function $Q$
\begin{equation}
E^{B}(S,\beta) =
Q\left(\frac{1}{\beta+1},\left(\Gamma(\frac{\beta+2}{\beta+1})S
\right)^{\beta+1}\right).
\label{eq:Gammafunc}
\end{equation}
Indeed this phenomenological two-parameter distribution provides an excellent
fit to our spectral data of 3,000 consecutive levels as can be seen in figure
5, with the best fit parameter values $\rho_1=0.309$ and $\beta =0.370$. They
differ slightly from the ideal values $\rho_1=0.365$ and $\beta=0.286\pm0.001$,
but are consistent with them thereby confirming the statistical independence
assumption.
\\\\
Nevertheless, just as a consistency check we can try to fit the spectrum with
the Brody distribution and the Berry-Robnik distribution. The result is shown
in figure 6 and the best fit parameters are $\beta=0.166\pm0.007$ and
$\rho_1=0.504$, respectively. The fits are not bad and Brody is certainly
better than Berry-Robnik as can be seen especially in the U-function plot.
The conclusion is that as a consequence of the localization effects the entire
spectrum is indeed best described by the Berry-Robnik-Brody distribution fitted
in figure 5. The value of $\chi^{2}$ in the latter is about three times smaller
than in Brody and six times smaller than in Berry-Robnik.
\\\\
Finally we should mention that the best fit Seligman-Verbaarschot
(\ref{eq:DSV2}) delta statistics fitted to the spectrum within the interval
$0.9<L<9.9$, below the saturation region, gives $\rho_1=0.472$ which is
consistent with the Berry-Robnik fit with $\rho_1=0.504$ as explained above.
For the sake of completeness we show the delta statistics in figure 7.
\\\\
Unlike the counting measure of regular levels the statistical measures of the
spectral subsequences do not converge so uniformly and so fast.
Calculating the level spacing distribution and the delta statistics of an
enlarged sample (namely 1,000 more levels added on top to the block of
3,000 levels) led actually to slightly worse results.
\\\\
Finally, as promised, we would like to give additional a posteriori reasons for
classifying the preliminary mixed type states as regular. (1) Due to the
strong localization any ambiguous mixed states is likely at least to mimic
a regular state. (2) After absorbing the mixed states into the set of regular
ones the relative fraction of regular states $\rho_1=0.365$ agrees much better
with the classical estimate $\rho_1=0.360$. (3) The statistical measures
($P(S)$
and $\Delta(L)$) also agree then much better with Poisson statistics. (4) The
statistical measures of irregular sequence agree then notably better with the
trend towards GOE in the sense that the Brody exponent $\beta$ increases from
$\beta=0.276$ to $\beta=0.286$.

\section{Discussions and conclusions}

The main goal of our present paper is to analyze a large sample of eigenstates
of an autonomous Hamiltonian system sufficiently high in the semiclassical
limit
and to perform the classification of the states into regular and irregular by
examining their Husimi-type phase space plots in correspondence with the
classical surface of section plots. This plan has materialized in the specific
choice of the Robnik billiard with the shape parameter $\lambda=0.15$, which is
a KAM-type system (the boundary is convex and analytic) with one large and
dominating chaotic component with the relative phase space volume (not SOS
area) $\rho_2=0.640$. We have plotted 3,000 consecutive states (from the
10,001st
to the 13,000th) in configuration and in phase space and performed the said
classification with the result that the relative fraction of regular states is
equal to $0.365$ which is in excellent agreement with the classical value
$\rho_1=0.360=1-\rho_2$.
Further we have calculated the statistical measures (level spacing distribution
and the delta statistics) for the regular and irregular sequence separately and
also of the entire spectrum. The aspects and assumptions of the Berry-Robnik
(1984) approach have been confirmed although we are not yet far enough in the
semiclassical limit to see the GOE statistics for the irregular levels, whilst
the regular levels are seen to obey the Poisson statistics very well. We
believe that this is the first direct dynamical separation of regular and
irregular levels in a generic system and the first such statistical analysis.
It confirms the Percival's (1973) classification scheme, the ingredients in the
Berry-Robnik theory (1984) and the specific rigorous results on convex
billiards by Lazutkin (1981,1991). It supports our claim that Berry-Robnik
(1984) theory is asymptotically exact theory. One of the future projects
should be in
going further into the semiclassical limit for which new methods are needed and
one such might be the employment of the quantum Poincar\'e mapping (Bogomolny
1992, Schanz and Smilansky 1994, Prosen 1994a,b,c,d). This might lead us to a
direct demonstration of the applicability of the Berry-Robnik theory with high
numerical and statistical significance which so far was successful only in an
abstract time-dependent Hamiltonian system, namely the compactified standard
map (Prosen and Robnik 1994a,b).

\section*{Acknowledgments}

We thank Toma\v z Prosen for a few computer programs and assistance in
using them. One of us (MR) acknowledges stimulating discussions with
Oriol Bohigas, Giulio Casati, Boris V. Chirikov, Frank Steiner and
Hans A. Weidenm\"uller.
The financial support by the Ministry of Science and
Technology of the Republic of Slovenia is gratefully acknowledged.

\vfill
\newpage
\section*{References}
B\"acker A and Steiner F 1994 {\em DESY preprint}\\\\
Berry M V 1977a {\em Phil. Trans. Roy. Soc. London} {\bf 287} 237\\\\
Berry M V 1977b {\em J. Phys. A: Math. Gen.} {\bf 10} 2083\\\\
Berry M V 1983 in  in {\em Chaotic Behaviour of Deterministic Systems
(Proc. NATO ASI Les Houches Summer School)} eds
Iooss G, Helleman R H G and Stora R (Amsterdam: Elsevier) p171\\\\
Berry M V 1985 {\em Proc. Roy. Soc. London} {\bf A400} 229\\\\
Berry M V 1989 {\em Proc. Roy. Soc. London} {\bf A423} 219\\\\
Berry M V and Robnik M 1984 {\em J. Phys. A: Math. Gen.} {\bf 17} 2413\\\\
Berry M V and Robnik M 1986 {\em J. Phys. A: Math. Gen.} {\bf 19} 649\\\\
Bohigas O, 1991  in {\em Chaos and Quantum Systems (Proc. NATO ASI Les Houches
Summer School)} eds M-J Giannoni, A Voros and J Zinn-Justin,
(Amsterdam: Elsevier) p87\\\\
Bohigas O, Tomsovic S and Ullmo 1990a {\em Phys. Rev. Lett.} {\bf 64} 1479\\\\
Bohigas O, Tomsovic S and Ullmo 1990b {\em Phys. Rev. Lett.} {\bf 65} 5\\\\
Bohigas O, Tomsovic S and Ullmo 1993 {\em Phys. Rep.} {\bf 223} 4\\\\
Bogomolny E B 1992 {\em Nonlinearity} {\bf 5} 805\\\\
Bruus H and Stone A D 1994 {\em Preprint} Dept. Phys. Yale University\\\\
Brody T A 1973 {\em Lett. Nuovo Cimento} {\bf 7} 482\\\\
Brody T A, Flores J , French J B, Mello P A, Pandey A and Wong S S M 1981 {\em
Rev. Mod. Phys.} {\bf 53} 385\\\\
Eckhardt B 1988 {\em Phys. Rep.} {\bf 163} 205\\\\
Frisk H 1990 Nordita {\em Preprint}.\\\\
Gaspard P, Rice S A, Mikeska H J and Nakamura K 1990 {\em Phys. Rev. A} {\bf
42} 4015\\\\
Giannoni M-J, Voros J and Zinn-Justin eds. 1991 {\em Chaos and Quantum Systems}
(North-Holland)\\\\
Gutzwiller M C 1990 {\em Chaos in Classical and Quantum Mechanics} (New York:
Springer)\\\\
Haake F 1991 {\em Quantum Signatures of Chaos} (Berlin: Springer).\\\\
Hayli A, Dumont T, Moulin-Ollagier J and Strelcyn J M 1987 {\em J. Phys. A:
Math. Gen} {\bf 20} 3237\\\\
Heller E J 1991  in {\em Chaos and Quantum Systems (Proc. NATO ASI Les Houches
Summer School)} eds M-J Giannoni, A Voros and J Zinn-Justin,
(Amsterdam: Elsevier) p547\\\\
Izrailev F M 1989 {\em J. Phys. A: Math. Gen.} {\bf 22} 865\\\\
Lazutkin V F 1981 {\em The Convex Billiard and the Eigenfunctions of
the Laplace Operator} (Leningrad: University Press) (in Russian)\\\\
Lazutkin V F 1991 {\em KAM Theory and Semiclassical Approximations to
Eigenfunctions} (Heidelberg: Springer Verlag)\\\\
Li Baowen and Robnik M 1994a {\em J. Phys. A: Math. Gen.} {\bf 27} 5509\\\\
Li Baowen and Robnik M 1994b {\em Preprint CAMTP/94-8}
submitted to {\em J. Phys. A: Math. Gen.} in October\\\\
Li Baowen and Robnik M 1994c to be submitted to {\em J. Phys. A: Math.
Gen.}\\\\
Li Baowen and Robnik M 1994d {\em Preprint CAMTP/94-11} to be submitted\\\\
Lopac V, Brant S and Paar V 1992 {\em Phys. Rev. A} {\bf 45} 3503\\\\
Lopac V, Brant S and Paar V 1994 submitted\\\\
Markarian R 1993 {\em Nonlinearity} {\bf 6} 819\\\\
Mehta M L 1991 {\em Random Matrices} (San Diego: Academic Press)\\\\
Percival I C 1973 {\em J. Phys. B: At. Mol. Phys.} {\bf 6} L229 \\\\
Press W H, Flannery B P, Teukolsky S A and Vetterling W T 1986 {\em Numerical
Recipes} (Cambridge: Cambridge University Press)\\\\
Prosen T and Robnik M 1993a {\em J. Phys. A: Math. Gen.} {\bf 26} 2371\\\\
Prosen T and Robnik M 1993b {\em J. Phys. A: Math. Gen.} {\bf 26} 5365\\\\
Prosen T and Robnik M 1994a {\em J. Phys. A: Math. Gen.} {\bf 27} L459\\\\
Prosen T and Robnik M 1994b {\em J. Phys. A: Math. Gen.}  {\em in press}\\\\
Prosen T 1994a  {\em J. Phys. A: Math. Gen.} {\bf 27} L709\\\\
Prosen T 1994b {\em Preprint CAMTP/94-3} sumbitted to {\em J. Math. Phys.} in
April\\\\
Prosen T 1994c {\em Preprint CAMTP/94-9} submitted to {\em
J. Phys. A: Math. Gen.} in November\\\\
Prosen T 1994d {\em Preprint CAMTP/94-12 (in preparation)}\\\\
Robnik M 1983 {\em J. Phys. A: Math. Gen.} {\bf 16} 3971\\\\
Robnik M 1984 {\em J. Phys. A: Math. Gen.} {\bf 17} 1049\\\\
Robnik M 1994 {\em J. Phys. Soc. Japan Suppl.} {\bf 63} 131\\\\
Robnik M and Berry M V 1986 {\em J. Phys. A. Math. Gen} {\bf 19} 669\\\\
Schanz H and Smilansky U 1993 {\em Preprint Weizmann Institute}\\\\
Seligman T H and Verbaarschot J J M  1985 {\em J. Phys. A: Math. Gen.} {\bf 18}
2227\\\\
Stone A D and Bruus H 1993 {\em Physica B} {\bf 189} 43\\\\
Stone A D and Bruus H 1994 {\em Surface Science} in press\\\\
Takami T and Hasegawa H 1992  {\em Phys. Rev. Lett.} {\bf 68} 419\\\\
von Oppen F 1994a {\em Phys. Rev. Lett.} {\bf 73} 798\\\\
von Oppen F 1994b {\em Phys. Rev. Lett.} to appear
(Private Communication from H. A.
Weidenm\"uller)\\\\
Zakrzewski J and Delande D 1993 {\em Phys. Rev. E} {\bf 47} 1650\\\\

\vfill
\newpage
\newpage
\section*{Figure captions}

\bigskip
\bigskip

\noindent
{\bf Figure 1:} The cumulative level spacing distribution (a) and
U-function (b) of 1096 regular levels. In (a) the dotted curves
represent the Poisson and GOE statistics. The thick curve is the numerical
data, while the thin curve shows the best fit Berry-Robnik distribution and the
broken curve shows the best fit Brody distribution. The best fit parameters
are $\rho_1=0.704$, $\beta=0.041\pm0.002$, respectively.
In (b) we display the U-function difference $U(W)-U(W^{BR})$ versus W(S), where
$U_{br}=U(W^{BR}(S))$ is the U-function of the best fit Berry-Robnik
distribution.
The numerical noisy curves (the average value with $\pm1\sigma$ band)
in (b) are compared with the best fit Brody distribution (broken
curve) and the Poisson distribution (dotted curve).

\bigskip \bigskip

\noindent {\bf Figure 2:}
The delta statistics of 1096 regular levels and the best fit
Seligman-Verbaarschot formula (\ref{eq:DSV2}). The dotted curves are the
Poisson
and GOE, the thick curve shows the numerical data, the best fit
Seligman-Verbaarschot formula is represented by the thin curve which overlaps
the limiting Poissonian at $L<7.5$. The vertical dashed lines
indicate the region where the least square fit has been done. The best fit
parameter value is $\rho_1=0.9998$ which is in excellent agreement with the
ideal value 1.
\bigskip
\bigskip

\noindent {\bf Figure 3:}
The same as figure 1 but for the 1904 irregular levels. The best fit parameter
values are $\rho_1=0.374$ and $\beta=0.286\pm0.001$ for the Berry-Robnik and
Brody distributions, respectively. In the U-function plot (b) we see that the
Brody fit is globally very good. (We do not show the Poissonian curve here.)
\bigskip
\bigskip

\noindent {\bf Figure 4:}
The same as figure 2 but for the 1904 irregular levels. The best fit parameter
value is $\rho_1=0.373$ which is at variance with the ideal value 0, but is
completely consistent with the Berry-Robnik fit in figure 3.

\bigskip
\bigskip

\noindent {\bf Figure 5:}
The cumulative level spacing distribution (a) and
U-function plot (b) of the total 3,000 regular and irregular levels
compared with
the best fit  Berry-Robnik-Brody formula (\ref{eq:EBRB}). In (a) the
dotted curves represent the Poisson and GOE statistics. The thick curve is
the numerical data, while the thin curve shows the best fit Berry-Robnik-Brody
distribution. The best fit parameters are
$\rho_1=0.309$ and $\beta=0.370$. In (b) we show the U-function difference
$U(W)-U(W^{BRB})$ versus $W(S)$, where $U_{brb}=U(W^{BRB}(S))$ is the
U-function of the Berry-Robnik-Brody.
The thick curve represents numerical data and  the noisy curves indicate the
$\pm1\sigma$ band. The quality of the fit is really excellent. However, the
ideal values of $\rho_1$ and $\beta$ would be $\rho_1=0.360$ and, according to
figure 3, $\beta=0.286$.

\bigskip
\bigskip
\noindent {\bf Figure 6:}
The same as figure 3 but for the total 3,000 levels. The best fit parameter
values are $\rho_1=0.504$, $\beta=0.166\pm0.007$.
\bigskip
\bigskip

\noindent {\bf Figure 7:}
The same as figure 2 but for the total 3,000 levels. The best fit
parameter value is $\rho_1=0.472$, which is in complete agreement with the
Berry-Robnik fit in figure 6.
\end{document}